# Strain-tunable Single Photon Sources in WSe$_2$ Monolayers

*Oliver Iff [1&], Davide Tedeschi [2&], Javier Martín-Sánchez [3,4&*], Magdalena Moczała-Dusanowska[1], Sefaattin Tongay [5], Kentaro Yumigeta[5], Javier Taboada-Gutiérrez [3,4], Matteo Savaresi[2], Armando Rastelli[6], Pablo Alonso-González [3,4], Sven Höfling [1,7], Rinaldo Trotta [2*] and Christian Schneider [1*]*

[1] *Technische Physik and Wilhelm Conrad Röntgen Research Center for Complex Material Systems, Physikalisches Institut, Universität Würzburg, Am Hubland, D-97074 Würzburg, Germany*

[2] *Department of Physics, Sapienza University of Rome, Piazzale A. Moro 5, 00185 Rome, Italy*

[3] *Department of Physics, University of Oviedo, Oviedo, Spain*

[4] *Center of Research on Nanomaterials and Nanotechnology, CINN (CSIC—Universidad de Oviedo), El Entrego 33940, Spain.*

[5] *Arizona State University, Glendale, Arizona, United States of America*

[6] *Institute of Semiconductor and Solid State Physics, Johannes Kepler University Linz, Altenbergerstraße 69, 4040, Linz, Austria*

[7] *SUPA, School of Physics and Astronomy, University of St. Andrews, St Andrews, KY16 9SS, United Kingdom*

[*] *Corresponding author: Christian.schneider@physik.uni-wuerzburg.de, rinaldo.trotta@uniroma1.it, javiermartin@uniovi.es*

[&] *These authors contributed equally to this work*

ABSTRACT.

The appearance of single photon sources in atomically thin semiconductors holds great promises for the development of a flexible and ultra-compact quantum technology, in which elastic strain engineering can be used to tailor their emission properties. Here, we show a compact and hybrid 2D-semiconductor-piezoelectric device that allows for controlling the energy of single photons emitted by quantum emitters localized in wrinkled $WSe_2$ monolayers. We demonstrate that strain fields exerted by the piezoelectric device can be used to tune the energy of localized excitons in $WSe_2$ up to 18 meV in a reversible manner, while leaving the single photon purity unaffected over a wide range. Interestingly, we find that the magnitude and in particular the sign of the energy shift as a function of stress is emitter dependent. With the help of finite element simulations we suggest a simple model that explains our experimental observations and, furthermore, discloses that the type of strain (tensile or compressive) experienced by the quantum emitters strongly depends on their localization across the wrinkles. Our findings are of strong relevance for the practical implementation of single photon devices based on two-dimensional materials as well as for understanding the effects of strain on their emission properties.

KEYWORDS: single photon emitters, 2D materials, elastic strain engineering, photoluminescence, tungsten diselenide monolayers, piezoelectric devices

MAIN TEXT

The family of two-dimensional (2D) semiconductor transition metal dichalcogenides (TMDs), including $WS_2$, $WSe_2$, $MoS_2$ or $MoSe_2$, offers several advantages for optoelectronic and photonic applications. They possess a variety of properties such as direct bandgap when thinned down to the monolayer, quantum confinement due to their reduced out-of-plane dimensionality, large oscillator strength and quantum efficiency, optically controlled injection of electrons with defined spins for quantum spintronics and spin-photon interfacing [1,2]. Moreover, functional multilayer heterostructures can be easily built up by simply piling up one material on top of the other – not possible by conventional epitaxial growth techniques – and ultra-compact optoelectronic devices such as light emitting diodes have been already fabricated[3]. Another major advantage of 2D materials compared to conventional semiconductors is their impressive "stretchability", as they can withstand strain magnitudes well above 1% before mechanical rupture takes place. This offers a large playground for elastic strain engineering, since externally applied strain – by direct bending or using piezoelectric actuators [4–9] – provides a natural strategy to tailor the electronic and optical properties of the material.

The discovery of single photon emitters (SPEs) in 2D materials has stimulated an intensive research effort, aimed at the exploitation of such sources for quantum photonics[10] as well as at the understanding of their physical origin. In the last years, SPEs have been reported on TMDs monolayers at cryogenic temperatures [11–15] and on layered hexagonal boron nitride at room temperature [16]. The origin of SPEs has been attributed to either presence of defects in the crystalline structure of the crystals[16,17] or bandgap modulation of the material due to local bending of the material itself, which naturally occurs on bubbles or wrinkles [18]. In an attempt to realize scalable SPEs over large areas, quantum dot-like nanostructures have been fabricated in atomically thin $WSe_2$ or $WS_2$ bilayers and monolayers by local strain engineering: Therein predefined areas of the monolayer were elastically deformed by pillars or nanorods[19–21]. A recent theoretical work has suggested that this deformation induces the confinement of excitons in potential wells with bound states that might lead to emission of single photons [18]. First attempts to couple $WSe_2$ quantum emitters to optical resonances of metallic nanostructures [22–25] and waveguide structures [26,27] have been also reported, which outlines the possibility for integration of TMD monolayers with optical cavities and on-chip integrated quantum photonic circuits.

In spite of these significant advances, quantum emitters in 2D materials deliver photons at random energies, which are difficult to control owing to the complexity either of the potential profile eventually leading to the exciton confinement or to the nature of the defects. In turn, this severely limits the suitability of 2D SPEs for applications in quantum information science and technology. Therefore, it is fundamental to develop post-fabrication tuning methods capable of controlling the SPEs emission energy in a reversible manner while leaving their optical quality unaffected. Elastic strain engineering of the material's band structure is a promising strategy to accomplish this task, as previously demonstrated on self-assembled III-V quantum dots[28,29]. Previous results on defect-induced emitters in thick hBN flakes reported moderate shift values of about 6 meV/%, likely due to the poor strain transfer efficiency of the substrate-bending technique used in the experiments[30]. To the best of our knowledge, no systematic and comprehensive studies on the strain effects, including the impact on the single photon purity in SPEs in TMDs has been published until now.

In this work, we successfully demonstrate active tuning of the emission energy of SPEs in WSe$_2$ monolayers using (001)- and (110)- [Pb(Mg1/3Nb2/3)O3]$_{0.72}$-[PbTiO3]$_{0.28}$ (PMN-PT) piezoelectric actuators. We show that a reversible tuning range of up to 18 meV can be obtained for moderate applied electric fields of about 15 kV/cm and that the single photon purity – as measured via the second order correlation function – is unaffected by strain, with a value g$^{(2)}$(0) bound to ~0.12 over a range of 5 meV. Interestingly, we find that upon the same applied voltage, different quantum emitters show completely different energy-shifts, both in magnitude and sign. Using finite element calculations, we discuss that this effect can be attributed to a different location of the quantum emitters across a wrinkle, which "feel" different strain fields (tensile and compressive) for the same induced strain field delivered by the piezo-actuator.

We first study the possibility of implementing a hybrid 2D-WSe$_2$-piezoelectric device capable of delivering single photons with controlled energy. Fig. 1a shows a sketch of our device: a wrinkled WSe$_2$ monolayer flake obtained by mechanical exfoliation is transferred on top of a PMN-PT piezoelectric plate coated with a [Cr (3 nm)/Au (100 nm)] bilayer on both sides for electrical contact [31]. In a final step, the full device is mounted onto an AlN chip carrier providing electrical contacts. The application of an electric field along the poling direction of the piezoelectric actuator (F$_p$) produces an out-of-plane deformation of the plate which induces an in-plane deformation to the attached monolayer, as sketched in figure 1a). In particular, a positive (negative) electric field leads to a compressive (tensile) in-plane deformation of the piezo-crystal which is transferred to the attached WSe$_2$ monolayer. Spatially resolved optical spectroscopy was performed in a micro-photoluminescence (μPL) setup. The samples were excited non-resonantly at 632.8 nm in an optical cryostat. Fig. 1b) shows a representative photoluminescence spectrum from a WSe$_2$ monolayer at a nominal sample temperature of 5 K. At elevated pump powers, the spectrum features a characteristic line-shape revealing luminescence from the free exciton (X) and trion (X-) resonances, as well as a broad, structured peak spanning over the spectral range from 730 nm up to 770 nm widely attributed to quantum emitters localized in strained areas related to bubbles or wrinkles [18,19,22,23]. Atomic force microscopy images of representative wrinkles in the monolayer are shown in Fig. S1 (Supporting Information). We could identify a variety of emission lines at low excitation power, which are spectrally well-isolated, and characterized by emission linewidth on the order of 200 μeV (see Fig. 1c)). We will refer hereafter to SPEs localized on wrinkled areas of the monolayer.

Spectral control over one of the observed quantum emitters is demonstrated by sweeping the electric field across to the piezoelectric actuator from $F_p$=-20 kV/cm to $F_p$=20 kV/cm as shown in Fig. 2a). Specifically, an emission energy blue/red shift is observed for compressive/tensile strain fields introduced by the actuator, which is attributed to an increase/decrease of the bandgap in crystalline semiconductor materials. Moreover, the shift of the emission can be reversibly tuned in a linear fashion, as in principle expected for moderate magnitudes of the strain delivered by the piezo-actuator (see below). From a linear fit to the measured spectra, we can infer a total energy shift of about 5 meV with a ratio 5.4 μeV/V, similar to the values reported in semiconductor nanomembranes containing quantum dots[32]. Based on previous experiments, we expect that for the maximum electric field investigated an in-plane biaxial strain value of about 0.15% [32]. It should be mentioned that, due to the Gaussian-like geometry of the wrinkle in the WSe$_2$ monolayer, the induced strain magnitude and sign strongly depends on the specific SPE under investigation, as discussed in detail below.

Whether single photons with high purity can be emitted at selected energies is further studied by second order correlation measurements. To do so, we investigate the influence of induced strain fields on the spectral shape and quality of the single photon emission as seen in Fig. 2. The luminescence is spectrally filtered by passing through a monochromator with 0.3 nm bandwidth and coupled to a fiber connected Hanbury Brown and Twiss (HBT) setup. The HBT set-up is equipped with two avalanche photodiodes (featuring time jitter of about 400 ps) connected to the correlation electronics. Notably, we observe a well-pronounced anti-bunching signal at zero delay times ($\tau = 0$), allowing us to extract a deconvoluted $g^{(2)}(0)$ value of ~0.13[33] at 0 $F_p$ =kV/cm and 0.12 at both positive and negative $F_p$ =20kV/cm. From the single exponential decay of the correlation functions carried out in the non-saturation regime of the exciton, we can furthermore estimate the spontaneous emission lifetimes of the exciton with a value of about 1.0 ns, which is on the faster end of previously reported values [11]. Importantly, both the exciton lifetime, as well as the single photon purity of the emitter are fully retained in the presence of mechanical stress, as evidenced by the comparative plot in Fig. 2c. This unambiguously confirms that strain does not alter the quality of the single photons emitted by quantum emitters in 2D materials – at least for the strain magnitudes and anisotropy investigated in this work.

As mentioned above, different SPEs lines show different shifts as the electric field is applied to the piezo-actuator. As an example, Fig. 3a) shows sharp emission lines originating from another wrinkle, which exhibit red and blue shifts as the electric field is varied from $F_p$=0 to 15 kV/cm. At a first sight, this is surprising since a blue shift of all the lines is expected for compressive strain delivered by the actuator. In order to explain this peculiar finding, it is first important to note that light is collected over a region with a diameter of ~1 µm and, therefore, the whole wrinkle is probed in the experiment. Second, previous reports [16,17,19] have shown that SPEs are likely related to the presence of point defects in the monolayer, whose radiative efficiency is enhanced by strain fields via the funnel effect. It is therefore reasonable to assume that the different lines originate from different regions of the wrinkle where the initial strain distribution and its variation with the externally-applied stress may be completely different. To support this hypothesis, we perform finite element simulations (FEM) of the strain field distribution in a WSe$_2$ flake featuring a representative Gaussian-shaped wrinkle subject to a deformation induced by a piezoelectric actuator. Details about the simulations can be found in the Supporting Information. Although each wrinkle is slightly different, we simulate a wrinkle with dimensions 100 nm (width) and 40 nm (height) and around 1.5 μm (length) in agreement with AFM measurements (see Fig. S1). The WSe$_2$ flake is simulated as a continuum film with thickness of 0.5 nm. Fig. 3b) shows the hydrostatic strain distribution map $\varepsilon_{xx}+ \varepsilon_{yy}+ \varepsilon_{zz}$ (directly connected to the energy shift via the materials' deformation potentials) for a specific electric field $F_p$=30 kV/cm applied on a (001)-PMN-PT piezoelectric actuator, which has an isotropic piezoelectric response – i.e. nominally isotropic biaxial in-plane strain fields are delivered by this kind of plates. It should be noted that the strain reported in panel b) is not the absolute strain configuration of the wrinkle, but its variation as the electric field on the piezoelectric actuator is changed from $F_p$=0 to $F_p$=30 kV/cm. Hence, this map represents only the induced strain field on the wrinkle and its initial pre-stress configuration, since unknown, is not taken into account in the FEM simulation. As expected, a uniform compressive strain distribution is obtained on the flat areas of the monolayer in contact with the piezoelectric actuator (e.g., point D in fig. 3b), whereas a highly non-uniform strain field distribution is obtained in the suspended wrinkle. Most interestingly, the sign of the strain (tensile or compressive) strongly depends on the specific location across the wrinkle with a gradual decrease of the magnitude for compressive strain as we move from the tail to the top (points A to C in Fig. 3b)). On the other hand, for the wrinkle geometry studied here, tensile strain is found at specific positions at the edges of the wrinkle. Moreover, also the slope strain/electric field is position dependent as shown in Fig. 3c) – with relatively small values for tensile and large values for compressive strain. While the comparison between theory and experiment would suggest that the lines red-shifting (blue-shifting) originate from the edges (body) of the wrinkle, this behavior strongly depends on the specific geometry of the wrinkle and/or the type of strain delivered by the piezoelectric actuator (see supporting information). To support this statement we performed additional experiments using a (110)-PMN-PT piezoelectric plate, which nominally provides highly anisotropic in-plane strain fields ($\varepsilon_{xx} \approx -0.3\varepsilon_{yy}$), where "x" and "y" directions correspond to [100] and [01-1] crystallographic directions of the PMN-PT crystal, respectively[34]. Interestingly, exactly the opposite behavior is observed in this case as shown in Fig. 4a, with a smaller blue-shift for the line at high energy and a remarkably large red-shift up to 18 meV for the lower energy lines (for a maximum electric field $F_p$=15 kV/cm applied to the actuator). It should be mentioned that due to the relatively high strain anisotropy introduced by (110)-oriented piezoelectric plates, a completely different strain distribution on the wrinkle is to be expected. This is demonstrated by FEM simulations in Fig. 4b, where the hydrostatic strain map for a Gaussian-shaped wrinkle oriented at 45 degrees with respect to the [100] direction of the piezoelectric plate is shown. We note that our experimental observations depicted in Fig. 4a are compatible with a situation where the emitters are located at the edge of a wrinkle (points A and B in Fig. 4b). Interestingly, in this case, a large/small tensile/compressive strain field is obtained which may explain our findings (Fig. 4c). We emphasize once more the importance of the initial geometry of the wrinkle and additional experiments are needed to clarify whether there is tight connection between strain-slope and position of the emitter. While we leave this point to future studies, it is worth emphasizing that our findings are compatible with a scenario of ref [17] (in which multiple SPEs are located at different points of the wrinkle) but not fully consistent with the idea that quantum emitters are solely located on the regions with a significant bending[18].

In summary, we have demonstrated active control of the energy emission of SPEs localized in a wrinkled $WSe_2$ monolayer. This is achieved developing a hybrid 2D-piezoelectric device where in-plane biaxial strain fields up to a magnitude of ~0.15% can be transferred to SPEs without degrading their optical quality. We demonstrate a record tunability up to 18 meV – much larger than the best reported values in 2D materials [30]. Moreover, the SPEs retain a high-purity single photon emission upon the introduction of strain fields, as shown by time-correlation measurements[28]. Finally, we have observed that different SPEs located in the same wrinkle shift to both higher and lower energies for the same applied stress. FEM simulations suggest that this peculiar behavior is related to the specific location of the SPEs and to the high non-uniform strain distribution across the wrinkle. The results reported in this work pave the way towards the exploitation of energy-tunable SPEs and emitters of entangled photon pairs[35] based on two dimensional crystals and it will stimulate the use of strain-fields to understand the origin of SPEs in 2D materials. For this last point, we anticipate that anisotropic strain fields delivered by micro-machined piezoelectric actuators [28,36] will have a key role where 2D materials can be incorporated upon integration in dielectric nanomembranes [37].

Acknowledgements:

J.M.-S. acknowledges support through the Clarín Programme from the Government of the Principality of Asturias and a Marie Curie-COFUND European grant (PA-18-ACB17-29). P.A.-G. acknowledges support from the European Research Council under Starting Grant 715496, 2DNANOPTICA. R. T., D. T. and M. S. acknowledge support by the European Research council (ERC) under the European Union's Horizon 2020 Research and Innovation Programme (SPQRel, Grant agreement No. 679183). C.S. acknowledges support by the European Research council (ERC) under the European Union's Horizon 2020 Research and Innovation Programme (UnLiMIt-2D), Grant agreement No. 679288). A.R. acknowledges support from the Linz Institute of Technology (LIT). The Würzburg group acknowledges support by the State of Bavaria. S.T acknowledges support from NSF DMR-1552220 and NSF DMR-1838443.

References:

(1)     Kobolov, K. S.; Tominaga, J. *Two-Dimensional Transition Metal Dichalcogenides*; Springer Series in Materials Science: Switzerland, 2016. https://doi.org/10.1007/978-3-319-31450-1_7.

(2) Liu, G. Bin; Pang, H.; Yao, Y.; Yao, W. Intervalley Coupling by Quantum Dot Confinement Potentials in Monolayer Transition Metal Dichalcogenides. *New J. Phys.* **2014**. https://doi.org/10.1088/1367-2630/16/10/105011.

(3) Withers, F.; Del Pozo-Zamudio, O.; Mishchenko, A.; Rooney, a. P.; Gholinia, A.; Watanabe, K.; Taniguchi, T.; Haigh, S. J.; Geim, a. K.; Tartakovskii, a. I.; et al. Light-Emitting Diodes by Band-Structure Engineering in van Der Waals Heterostructures. *Nat. Mater.* **2015**, *14* (February), 301–306. https://doi.org/10.1038/nmat4205.

(4) Castellanos-Gomez, A.; Roldán, R.; Cappelluti, E.; Buscema, M.; Guinea, F.; Van Der Zant, H. S. J.; Steele, G. A. Local Strain Engineering in Atomically Thin MoS2. *Nano Lett.* **2013**. https://doi.org/10.1021/nl402875m.

(5) Desai, S. B.; Seol, G.; Kang, J. S.; Fang, H.; Battaglia, C.; Kapadia, R.; Ager, J. W.; Guo, J.; Javey, A. Strain-Induced Indirect to Direct Bandgap Transition in Multilayer WSe 2. *Nano Lett.* **2014**, *14* (8), 4592–4597. https://doi.org/10.1021/nl501638a.

(6) Martín-Sánchez, J.; Trotta, R.; Mariscal, A.; Serna, R.; Piredda, G.; Stroj, S.; Edlinger, J.; Schimpf, C.; Aberl, J.; Lettner, T.; et al. Strain-Tuning of the Optical Properties of Semiconductor Nanomaterials by Integration onto Piezoelectric Actuators. *Semicond. Sci. Technol.* **2018**, *33* (1). https://doi.org/10.1088/1361-6641/aa9b53.

(7) Island, J. O.; Kuc, A.; Diependaal, E. H.; Bratschitsch, R.; Van Der Zant, H. S. J.; Heine, T.; Castellanos-Gomez, A. Precise and Reversible Band Gap Tuning in Single-Layer MoSe2 by Uniaxial Strain. *Nanoscale* **2016**, *8* (5), 2589–2593. https://doi.org/10.1039/c5nr08219f.

(8) Niehues, I.; Schmidt, R.; Drüppel, M.; Marauhn, P.; Christiansen, D.; Selig, M.; Berghäuser, G.; Wigger, D.; Schneider, R.; Braasch, L.; et al. Strain Control of Exciton–Phonon Coupling in Atomically Thin Semiconductors. *Nano Lett.* **2018**, *18* (3), 1751–1757. https://doi.org/10.1021/acs.nanolett.7b04868.

(9) Kumar, S.; Kaczmarczyk, A.; Gerardot, B. D. Strain-Induced Spatial and Spectral Isolation of Quantum Emitters in Mono- and Bilayer WSe2. *Nano Lett.* **2015**, *15* (11), 7567–7573. https://doi.org/10.1021/acs.nanolett.5b03312.

(10) Aharonovich, I.; Englund, D.; Toth, M. Solid-State Single-Photon Emitters. *Nat. Photonics* **2016**, *10* (10), 631–641. https://doi.org/10.1038/nphoton.2016.186.

(11) Tonndorf, P.; Schmidt, R.; Schneider, R.; Kern, J.; Buscema, M.; Steele, G. a.; Castellanos-Gomez, A.; van der Zant, H. S. J.; Michaelis de Vasconcellos, S.; Bratschitsch, R. Single-Photon Emission from Localized Excitons in an Atomically Thin Semiconductor. *Optica* **2015**, *2* (4), 347. https://doi.org/10.1364/OPTICA.2.000347.

(12) Srivastava, A.; Sidler, M.; Allain, A. V.; Lembke, D. S.; Kis, A.; Imamoğlu, A. Optically Active Quantum Dots in Monolayer WSe2. *Nat. Nanotechnol.* **2015**, *10* (6), 491–496. https://doi.org/10.1038/nnano.2015.60.

(13) He, Y.-M.; Clark, G.; Schaibley, J. R.; He, Y.; Chen, M.-C.; Wei, Y.-J.; Ding, X.; Zhang, Q.; Yao, W.; Xu, X.; et al. Single Quantum Emitters in Monolayer Semiconductors. *Nat. Nanotechnol.* **2015**, *10* (6), 497–502. https://doi.org/10.1038/nnano.2015.75.

(14) Koperski, M.; Nogajewski, K.; Arora, A.; Cherkez, V.; Mallet, P.; Veuillen, J.-Y.; Marcus, J.; Kossacki, P.; Potemski, M. Single Photon Emitters in Exfoliated WSe2 Structures. *Nat.*

*Nanotechnol.* **2015**, *10* (6), 503–506. https://doi.org/10.1038/nnano.2015.67.

(15) Chakraborty, C.; Kinnischtzke, L.; Goodfellow, K. M.; Beams, R.; Vamivakas, a. N. Voltage-Controlled Quantum Light from an Atomically Thin Semiconductor. *Nat. Nanotechnol.* **2015**, *10* (6), 507–511. https://doi.org/10.1038/nnano.2015.79.

(16) Tran, T. T.; Bray, K.; Ford, M. J.; Toth, M.; Aharonovich, I. Quantum Emission from Hexagonal Boron Nitride Monolayers. *Nat. Nanotechnol.* **2016**, *11* (1), 37–41. https://doi.org/10.1038/nnano.2015.242.

(17) Zheng, Y. J.; Chen, Y.; Huang, Y. L.; Gogoi, P. K.; Li, M.-Y.; Li, L.-J.; Trevisanutto, P. E.; Wang, Q.; Pennycook, S. J.; Wee, A. T. S.; et al. Point Defects and Localized Excitons in 2D WSe 2. *ACS Nano* **2019**, *13* (5), 6050–6059. https://doi.org/10.1021/acsnano.9b02316.

(18) Chirolli, L.; Prada, E.; Guinea, F.; Roldán, R.; San-Jose, P. Strain-Induced Bound States in Transition-Metal Dichalcogenide Bubbles. *2D Mater.* **2019**, *6* (2), 25010. https://doi.org/10.1088/2053-1583/ab0113.

(19) Branny, A.; Kumar, S.; Proux, R.; Gerardot, B. D. Deterministic Strain-Induced Arrays of Quantum Emitters in a Two-Dimensional Semiconductor. *Nat. Commun.* **2017**, *8* (May), 1–7. https://doi.org/10.1038/ncomms15053.

(20) Palacios-Berraquero, C.; Kara, D. M.; Montblanch, A. R. P.; Barbone, M.; Latawiec, P.; Yoon, D.; Ott, A. K.; Loncar, M.; Ferrari, A. C.; Atatüre, M. Large-Scale Quantum-Emitter Arrays in Atomically Thin Semiconductors. *Nat. Commun.* **2017**, *8* (May), 1–6. https://doi.org/10.1038/ncomms15093.

(21) Kern, J.; Niehues, I.; Tonndorf, P.; Schmidt, R.; Wigger, D.; Schneider, R.; Stiehm, T.;

Michaelis de Vasconcellos, S.; Reiter, D. E.; Kuhn, T.; et al. Nanoscale Positioning of Single-Photon Emitters in Atomically Thin WSe2. *Adv. Mater.* **2016**, 7101–7105. https://doi.org/10.1002/adma.201600560.

(22) Tripathi, L. N.; Iff, O.; Betzold, S.; Emmerling, M.; Moon, K.; Lee, Y. J.; Kwon, S.-H.; Höfling, S.; Schneider, C. Spontaneous Emission Enhancement in Strain-Induced WSe2 Monolayer Based Quantum Light Sources on Metallic Surfaces. *ACS Photonics* **2018**, *5* (5), 1919–1926. https://doi.org/10.1021/acsphotonics.7b01053.

(23) Iff, O.; Lundt, N.; Betzold, S.; Tripathi, L. N.; Emmerling, M.; Tongay, S.; Lee, Y. J.; Kwon, S.-H.; Höfling, S.; Schneider, C. Deterministic Coupling of Quantum Emitters in WSe 2 Monolayers to Plasmonic Nanocavities. *Opt. Express* **2018**, *26* (20), 25944. https://doi.org/10.1364/oe.26.025944.

(24) Cai, T.; Kim, J.; Yang, Z.; Dutta, S.; Aghaeimeibodi, S.; Waks, E. Radiative Enhancement of Single Quantum Emitters in WSe2 Monolayers Using Site-Controlled Metallic Nanopillars. *ACS Photonics* **2018**, *5*, 3466–3471. https://doi.org/10.1021/acsphotonics.8b00580.

(25) Luo, Y.; Shepard, G. D.; Ardelean, J. V; Rhodes, D. A.; Kim, B.; Barmak, K.; Hone, J. C.; Strauf, S. Deterministic Coupling of Site-Controlled Quantum Emitters in Monolayer WSe2 to Plasmonic Nanocavities. *Nat. Nanotechnol.* **2018**, *13* (December). https://doi.org/10.1038/s41565-018-0275-z.

(26) Chen, H.; Nanz, S.; Abass, A.; Yan, J.; Gao, T.; Choi, D.-Y.; Kivshar, Y. S.; Rockstuhl, C.; Neshev, D. N. Enhanced Directional Emission from Monolayer WSe2 Integrated onto a Multiresonant Silicon-Based Photonic Structure. *ACS Photonics* **2017**,

acsphotonics.7b00550. https://doi.org/10.1021/acsphotonics.7b00550.

(27) Hartwig, O.; Kaniber, M.; Finley, J. J.; Prechtl, M.; Cerne, J.; Vest, G.; Jürgensen, M.; Blauth, M. Coupling Single Photons from Discrete Quantum Emitters in WSe 2 to Lithographically Defined Plasmonic Slot Waveguides . *Nano Lett.* **2018**, *18* (11), 6812–6819. https://doi.org/10.1021/acs.nanolett.8b02687.

(28) Trotta, R.; Martín-Sánchez, J.; Wildmann, J. S.; Piredda, G.; Reindl, M.; Schimpf, C.; Zallo, E.; Stroj, S.; Edlinger, J.; Rastelli, A. Wavelength-Tunable Sources of Entangled Photons Interfaced with Atomic Vapours. *Nat. Commun.* **2016**, *7*. https://doi.org/10.1038/ncomms10375.

(29) Huang, H.; Trotta, R.; Huo, Y.; Lettner, T.; Wildmann, J. S.; Martín-Sánchez, J.; Huber, D.; Reindl, M.; Zhang, J.; Zallo, E.; et al. Electrically-Pumped Wavelength-Tunable GaAs Quantum Dots Interfaced with Rubidium Atoms. *ACS Photonics* **2017**, *4* (4), 868–872. https://doi.org/10.1021/acsphotonics.6b00935.

(30) Grosso, G.; Moon, H.; Lienhard, B.; Ali, S.; Efetov, D. K.; Furchi, M. M.; Jarillo-Herrero, P.; Ford, M. J.; Aharonovich, I.; Englund, D. Tunable and High-Purity Room Temperature Single-Photon Emission from Atomic Defects in Hexagonal Boron Nitride. *Nat. Commun.* **2017**, *8* (1), 705. https://doi.org/10.1038/s41467-017-00810-2.

(31) Castellanos-Gomez, A.; Buscema, M.; Molenaar, R.; Singh, V.; Janssen, L.; van der Zant, H. S. J.; Steele, G. a. Deterministic Transfer of Two-Dimensional Materials by All-Dry Viscoelastic Stamping. *2D Mater.* **2014**, *1* (1), 011002. https://doi.org/10.1088/2053-1583/1/1/011002.

(32) Trotta, R.; Atkinson, P.; Plumhof, J. D.; Zallo, E.; Rezaev, R. O.; Kumar, S.; Baunack, S.; Schröter, J. R.; Rastelli, A.; Schmidt, O. G. Nanomembrane Quantum-Light-Emitting Diodes Integrated onto Piezoelectric Actuators. *Adv. Mater.* **2012**, *24* (20), 2668–2672. https://doi.org/10.1002/adma.201200537.

(33) Regelman, D. V.; Mizrahi, U.; Gershoni, D.; Ehrenfreund, E.; Schoenfeld, W. V.; Petroff, P. M. Semiconductor Quantum Dot: A Quantum Light Source of Multicolor Photons with Tunable Statistics. *Phys. Rev. Lett.* **2001**. https://doi.org/10.1103/PhysRevLett.87.257401.

(34) Zhang, J.; Wildmann, J. S.; Ding, F.; Trotta, R.; Huo, Y.; Zallo, E.; Huber, D.; Rastelli, A.; Schmidt, O. G. High Yield and Ultrafast Sources of Electrically Triggered Entangled-Photon Pairs Based on Strain-Tunable Quantum Dots. *Nat. Commun.* **2015**, *6*, 1–7. https://doi.org/10.1038/ncomms10067.

(35) He, Y.-M.; Iff, O.; Lundt, N.; Baumann, V.; Davanco, M.; Srinivasan, K.; Hofling, S.; Schneider, C. Cascaded Emission of Single Photons from the Biexciton in Monolayered $WSe_2$. *Nat. Commun.* **2016**, In Press. https://doi.org/10.1038/ncomms13409.

(36) Martín-Sánchez, J.; Trotta, R.; Piredda, G.; Schimpf, C.; Trevisi, G.; Seravalli, L.; Frigeri, P.; Stroj, S.; Lettner, T.; Reindl, M.; et al. Reversible Control of In-Plane Elastic Stress Tensor in Nanomembranes. *Adv. Opt. Mater.* **2016**, *4* (5). https://doi.org/10.1002/adom.201500779.

(37) Martín-Sánchez, J.; Mariscal, A.; De Luca, M.; Martín-Luengo, A. T.; Gramse, G.; Halilovic, A.; Serna, R.; Bonanni, A.; Zardo, I.; Trotta, R.; et al. Effects of Dielectric Stoichiometry on the Photoluminescence Properties of Encapsulated WSe2 Monolayers. *Nano Res.* **2017**. https://doi.org/https://doi.org/10.1007/s12274-017-1755-4.

**Figures**

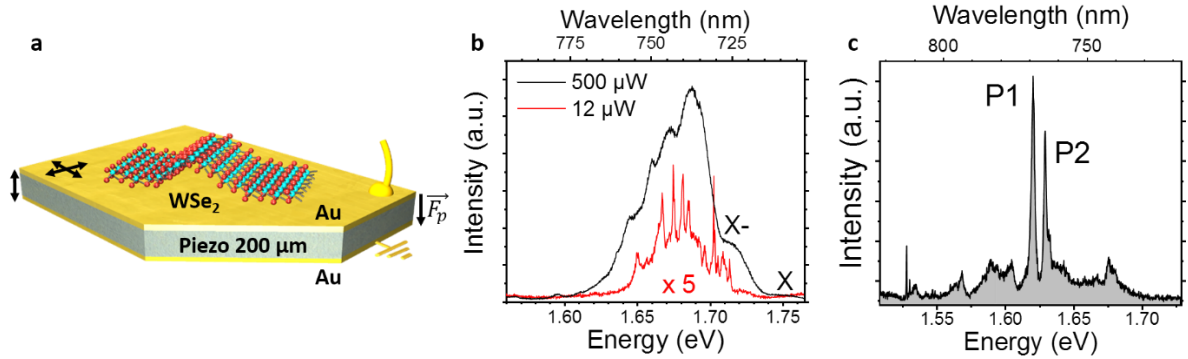

**Fig 1.** a) Schematics of a hybrid 2D-semiconductor-piezoelectric actuator with an integrated WSe$_2$ monolayer. b) Typical micro-photoluminescence spectrum of a WSe$_2$ monolayer, including the exciton (X) and the trion (X-), which breaks up at low pump power into discrete emission lines in the spectral range around 725-780 nm. c) Close up spectrum of a pair of selected lines P1 and P2 with a linewidth of 230 µeV.

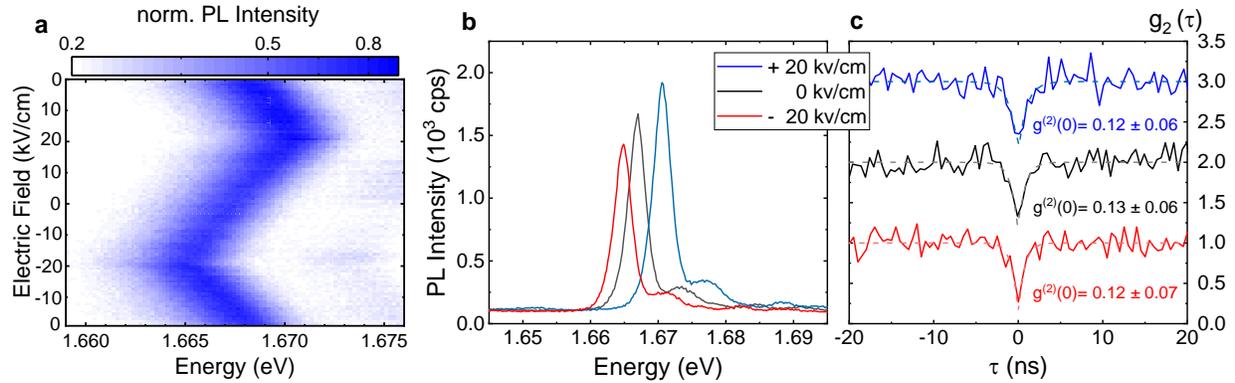

**Fig. 2** a) Contour plot of the µPL spectra of a SPE as a function of the applied electric field on the piezoelectric actuator. The electric field is reversibly swept in the range from -20 to 20 kV/cm. The observed red- and blue-shifts are due to the induced compressive and tensile strain fields by the actuator. An energy shift equal to 5.4 µeV/V is observed. b) PL spectra of the SPE in a) for zero and the maximum electric fields applied on the actuator (red and blue lines). A total shift of 5.4 meV is obtained for a total 40 kV/cm sweep. c) Second order auto-correlation $g^{(2)}(\tau)$ measurements of the dot in a) for $F_P$=-20, 0 and 20 kV/cm. The single photon emitter nature of the SPE is confirmed
16

by the low deconvoluted value $g^{(2)}(0)$~0.12, 0.13 and 0.12, respectively, which is independent of the applied field (i.e. induced strain field).

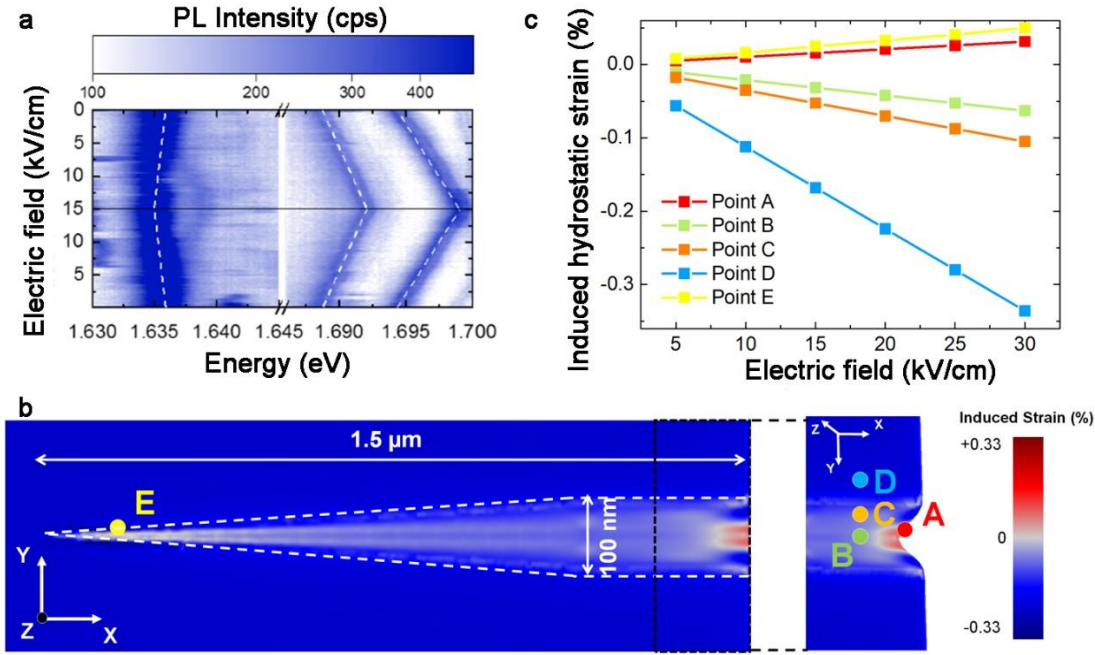

Fig. 3 a) Contour plot of the PL spectra of SPEs as a function of the electric field applied on a (001) piezoelectric device. The electric field was set up to $F_p$=15 kV/cm and back to $F_p$= 0 kV/cm. The black line in the middle highlights the point where the maximum voltage is applied. Three different peaks can be followed along the voltage sweep, two of which show a large blue-shift (6 meV), while the other one exhibits a small red-shift (1 meV). It is worth noting the absence of any kind of hysteresis along the voltage cycle on all the energy shifts. b) FEM simulation of the variation of the hydrostatic strain exerted over the wrinkle on a (001) piezoelectric actuator, biased at an electric field of $F_p$=30 kV/cm. A Gaussian profile with an aspect ratio of 2.5 are assumed as geometrical shape for the wrinkle. Both compressive and tensile strain fields are observed at different positions on the wrinkle upon application of the electric field to the piezoelectric device. c) Evolution of the hydrostatic strain as a function of the electric field applied to a (001) piezoelectric device calculated in 5 different points of the wrinkle, highlighted also in panel b). Independently of the type of strain, a linear dependence with the applied voltage is observed.

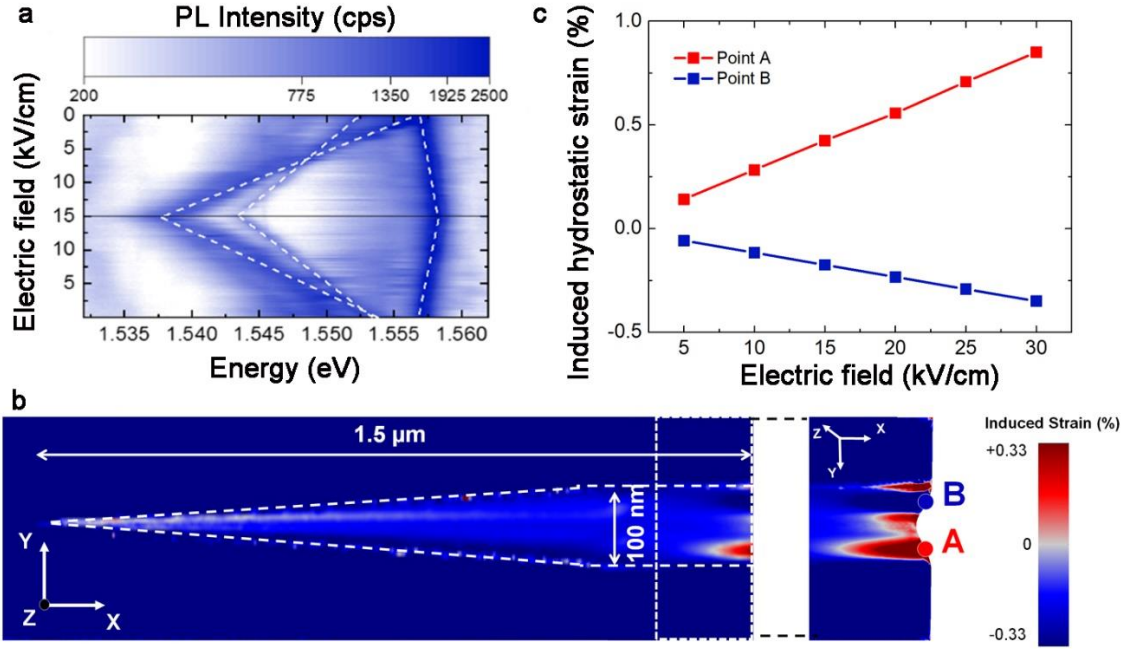

Fig. 4 a) Contour plot of the PL spectra of SPEs as a function of the electric field applied on a (110) piezoelectric device. The electric field was set up to $F_p=15$ kV/cm and back to $F_p= 0$ kV/cm. The black line in the middle highlights the point where the maximum voltage is applied. Three different peaks can be followed along the voltage sweep, among which two show a very large red-shift (up to 18 meV), while the other one exhibits a small red-shift (1 meV). b) FEM simulation of the variation of the hydrostatic strain exerted over the wrinkle on a (110) piezoelectric actuator, biased at an electric field of $F_p=30$ kV/cm. A Gaussian profile with an aspect ratio of 2.5 is assumed as geometrical shape for the wrinkle. Both compressive and tensile strain fields are observed at different positions on the wrinkle upon application of the electric field to the piezoelectric device. c) Evolution of the hydrostatic strain as a function of the electric field applied to the piezoelectric device calculated in 2 different points of the wrinkle, highlighted also in panel b).